\documentclass[conference]{IEEEtran}
\IEEEoverridecommandlockouts

\usepackage{cite}
\usepackage{amsmath,amssymb,amsfonts}
\usepackage{algorithmic}
\usepackage{graphicx}
\usepackage{textcomp}
\usepackage{xcolor}

\usepackage{lipsum}

\usepackage{graphicx}
\usepackage{tabularx}
\usepackage{enumitem}
\usepackage{MnSymbol}
\usepackage{wasysym}
\usepackage{pifont}
\usepackage{multirow}
\usepackage{caption}
\usepackage{subcaption}
\usepackage{titlesec}

\usepackage{hyperref}
\usepackage{array}
\usepackage{amsmath}
\usepackage{makecell}
\usepackage{colortbl}
\usepackage{soul}
\usepackage{amsmath}
\usepackage[ruled,vlined,linesnumbered]{algorithm2e}

\usepackage{booktabs}
\usepackage{layout}

\usepackage[top=0.7in,bottom=0.7in,left=0.58in,right=0.58in]{geometry}
\let\oldding\ding
\renewcommand{\ding}[2][1]{\scalebox{#1}{\oldding{#2}}}

\def\BibTeX{{\rm B\kern-.05em{\sc i\kern-.025em b}\kern-.08em
    T\kern-.1667em\lower.7ex\hbox{E}\kern-.125emX}}

\begin{document}

\title{\huge MIREDO: MIP-Driven Resource-Efficient Dataflow Optimization for Computing-in-Memory Accelerator}

\author{\IEEEauthorblockN{
Xiaolin He\IEEEauthorrefmark{2},
Cenlin Duan\IEEEauthorrefmark{3},
Yingjie Qi\IEEEauthorrefmark{2},
Xiao Ma\IEEEauthorrefmark{2},
Jianlei Yang\IEEEauthorrefmark{2}\IEEEauthorrefmark{4}
}
\IEEEauthorblockA{\IEEEauthorrefmark{2} School of Computer Science and Engineering, Beihang University}
\IEEEauthorblockA{\IEEEauthorrefmark{3} School of Integrated Circuit Science and Engineering, Beihang University}
\IEEEauthorblockA{\IEEEauthorrefmark{4} Qingdao Research Institute, Beihang University}
\thanks{This work is supported in part by the National Natural Science Foundation of China (Grant No. 62572036), the Beijing Natural Science Foundation (Grant No. L243031), the National Key R\&D Program of China (Grant No. 2023YFB4503704 and 2024YFB4505601).}
\thanks{Corresponding author is Jianlei Yang. Email: \href{mailto:jianlei@buaa.edu.cn}{jianlei@buaa.edu.cn}.}
}

\maketitle

\bstctlcite{IEEEexample:BSTcontrol}

\begin{abstract}

Computing-in-Memory (CIM) architectures have emerged as a promising solution for accelerating Deep Neural Networks (DNNs) by mitigating data movement bottlenecks. 
However, realizing the potential of CIM requires specialized dataflow optimizations, which are challenged by an expansive design space and strict architectural constraints. 
Existing optimization approaches often fail to fully exploit CIM accelerators, leading to noticeable gaps between theoretical and actual system-level efficiency.
To address these limitations, we propose the MIREDO framework, which formulates dataflow optimization as a Mixed-Integer Programming (MIP) problem. 
MIREDO introduces a hierarchical hardware abstraction coupled with an analytical latency model designed to accurately reflect the complex data transfer behaviors within CIM systems. 
By jointly modeling workload characteristics, dataflow strategies, and CIM-specific constraints, MIREDO systematically navigates the vast design space to determine the optimal dataflow configurations.
Evaluation results demonstrate that MIREDO significantly enhances performance, achieving up to $3.2\times$ improvement across various DNN models and hardware setups.

\end{abstract}

\section{Introduction}

Deep neural networks (DNNs) have achieved state-of-the-art (SOTA) performance across various domains, establishing themselves as the cornerstone of modern intelligent systems\cite{kwon2019understanding,he2016deep}.
However, the deployment of DNNs in resource-constrained edge devices faces significant challenges due to substantial computational and memory requirements. 
To reduce the memory bottlenecks and data transfer overhead inherent in traditional von Neumann architectures, computing-in-memory (CIM) has emerged as a promising paradigm~\cite{Shafiee2016ISAACAC}. 
By performing matrix-vector multiplications (MVMs) directly in-situ within memory arrays, CIM can significantly reduce costly data movement.
Prior works~\cite{houshmand2022diana,chang202240nm} have demonstrated the effectiveness of this paradigm, greatly advancing the field of specialized accelerators for DNN inference.

However, the efficient design of CIM accelerators faces unique challenges in dataflow optimization, which involves strategies for data partitioning and scheduling to optimize performance and energy efficiency. 
In contrast to traditional accelerators, dataflow optimization for CIM accelerators must consider additional factors: inherent row- and column-level parallelism constraints within CIM arrays, the bit-serial execution model, and the stalling introduced by weight reloading procedures~\cite{yang2023three}.
Empirically tuned or manually crafted mapping strategies often fail to fully exploit the potential of CIM architectures, resulting in substantial performance degradation.
Moreover, a single layer can have up to 13 million feasible dataflow options with drastically different performance characteristics~\cite{Park2023NeuroSpectorSO}, making exhaustive search computationally infeasible. 

These optimization issues often lead to a commonly observed gap between the theoretical performance of CIM macros and actual system-level performance, which primarily stems from mismatches between dataflow strategies and hardware architecture.
The challenge becomes more complex when considering the diversity of DNN workloads, each with distinct computational patterns and memory access requirements. 
Furthermore, the memory hierarchy external to the CIM macro introduces system-level bottlenecks that are heightened by limited on-chip resources~\cite{sun2023survey}. 
Therefore, optimizing the entire processing system, rather than individual macro-level design, is essential for realizing the full potential of CIM accelerators~\cite{Qi2025CIMFlowAI,houshmand2020opportunities}.
However, existing research offers limited insight into maximizing the system-level performance of CIM, particularly within resource-constrained accelerators. 
To bridge this gap, we introduce MIREDO, a dataflow optimization framework for SRAM-based CIM accelerators.
MIREDO employs Mixed-Integer Programming (MIP) for optimization, with a focus on data transfer efficiency.

The main contributions of this work are summarized as follows:
\begin{itemize}[leftmargin=20pt, topsep=0pt]
\item We present a hierarchical abstraction of CIM architectures and a corresponding data transfer analysis that reveals system-level performance bottlenecks.
\item We formulate the dataflow optimization as an MIP problem, which leverages a novel analytical latency model to guide the search for high-efficiency solutions.
\item We demonstrate that MIREDO achieves up to $3.2\times$ Energy-Delay Product (EDP) reduction across diverse workloads and hardware configurations.
\end{itemize}

\section{Background and Motivations}
This section provides an overview of CIM architecture and reviews the landscape of related work in dataflow design and optimization.
\begin{table}[t]
    \small
    \centering
        \caption{Comparison of works on different criteria.}
    \setlength{\tabcolsep}{10pt} 
    \renewcommand{\arraystretch}{1.2} 
    \resizebox{1\columnwidth}{!}{
        \begin{tabular}{cccc}
            \hline
            {\large \textbf{Work}} & \textbf{\makecell{System\\Optimization}} &  \textbf{\makecell{CIM\\Modeling}} &  \textbf{\makecell{CIM-aware\\Optimization}} \\
            \hline
            NeuroSpector\cite{Park2023NeuroSpectorSO} / CoSA\cite{huang2021cosa} & \ding{51} & \textbf{--} & \textbf{--} \\
            SPCIM\cite{wang2022spcim}  & \textbf{--} & \ding{51} & \ding{51} \\
            ZigZag-IMC\cite{sun2023analog}/CiMLoop\cite{andrulis2024cimloop}  & \ding{51} & \ding{51} & \textbf{--} \\
            \hline
            \textbf{MIREDO} & \ding{51} & \ding{51} & \ding{51} \\
            \hline
        \end{tabular}
    }
    \label{tab:comparison}
\end{table}

\subsection{CIM Architecture} 
CIM architectures represent a paradigm shift from conventional accelerator designs by performing computation directly within memory arrays, which mitigates the data-transfer bottleneck, thus significantly enhancing inference efficiency.
Previous studies~\cite{DBLP:journals/tcasI/MengWZL24, duan2023ddc, wang2020tcim} have shown that various technologies, such as RRAM, SRAM, and MRAM, are viable candidates for CIM accelerators.
Among these technologies, SRAM is particularly attractive due to its lower latency, higher endurance, and compatibility with advanced CMOS logic processes~\cite{Duan2024TowardsES,su2021two,Duan2025EfficientSC}.
While SRAM-based CIM architectures promise substantial performance gains for DNN inference, the inherent $6$T cross-coupled structure of SRAM results in much lower integration density and capacity than other technologies. 
As the scale of DNN models increases dramatically, multi-core CIM architectures have been widely explored to reduce the data transfer overhead and increase computation parallelism.
Although this approach improves execution efficiency, it introduces a series of system-level design complexities in memory mapping, data scheduling, and inter-core communication.
Therefore, developing a dataflow optimization strategy tailored for these multi-core systems has become critical to unleash their full potential.

\subsection{Dataflow Strategies}
Most existing CIM architectures leverage a weight-stationary (WS) dataflow, as it aligns well with in-situ computation principles~\cite{sun2023pimcomp}.
However, relying on such a static dataflow can lead to poor performance due to the aforementioned challenges. 
This highlights the need for adaptive optimization methods capable of generating dataflows tailored to diverse hardware configurations and the heterogeneity across different DNN layers.
A summary of several recent works addressing this problem is presented in Table~\ref{tab:comparison}.

While many dataflow optimization methods exist for traditional Processing Element (PE) accelerators, they fall short of leveraging the efficiency benefits of CIM.
For instance, NeuroSpector\cite{Park2023NeuroSpectorSO} is an analysis framework that highlights the impact of data movement on performance by modeling the memory hierarchy. 
CoSA\cite{huang2021cosa} employs MIP to model and optimize factors such as buffer utilization and communication traffic, a technique also employed in this paper.
The fundamental architectural differences between PE and CIM accelerators limit the effective application of these existing methods to CIM-based systems.

Among CIM-focused approaches, SPCIM~\cite{wang2022spcim} proposes a novel approach by introducing a reconfigurable cluster topology within CIM macro structures, which enables harnessing input parallelism or weight parallelism adaptively. 
However, the effectiveness of these optimizations is tightly coupled to a specific hardware topology, making it challenging to adapt efficiently to novel workloads and different hardware platforms.
ZigZag-IMC~\cite{sun2023analog} and CiMLoop~\cite{andrulis2024cimloop} provide comprehensive modeling and evaluation frameworks, effectively integrated into existing system-level infrastructures. 
These frameworks utilize random or heuristic search methods to find optimized dataflow configurations.

Although the aforementioned studies contribute valuable insights to specific aspects of CIM optimization, they exhibit the following limitations:

\raisebox{-0.1em}{\ding[1.3]{182}} \textbf{Oversimplified Performance Modeling:} 
Most existing works rely on simplistic performance models that assume data transfer latencies can be perfectly hidden through double-buffering~\cite{dave2019dmazerunner,russo2023memory}. 
These idealistic assumptions fail to capture the complex interactions within CIM architectures, leading to temporal underutilization of hardware resources and suboptimal system-level performance~\cite{mei2022uniform}.

\raisebox{-0.1em}{\ding[1.3]{183}} \textbf{Inefficient Design Space Exploration:}
Manual and heuristic methods depend heavily on predefined rules or expert knowledge, which lack scalability and cannot guarantee optimal solutions. 
Meanwhile, brute-force search becomes computationally prohibitive and is typically restricted to narrow, predefined search spaces.

To address these limitations, this work introduces MIREDO, an optimization framework that formulates dataflow optimization as an MIP problem. 
Based on a comprehensive analysis of CIM architectural characteristics and data transfer behaviors, our model incorporates resource constraints to accurately predict inference latency. 
This approach enables the systematic generation of high-performance dataflows for complex, multi-core CIM accelerators.

\section{Architecture Abstraction}\label{sec3}

To support the proposed MIREDO framework, we introduce a multi-core CIM architecture abstraction, characterizing its inherent structural parallelism and unique dataflow behaviors.      

\begin{figure}[t]
  \centering
  \includegraphics[width=0.95\columnwidth]{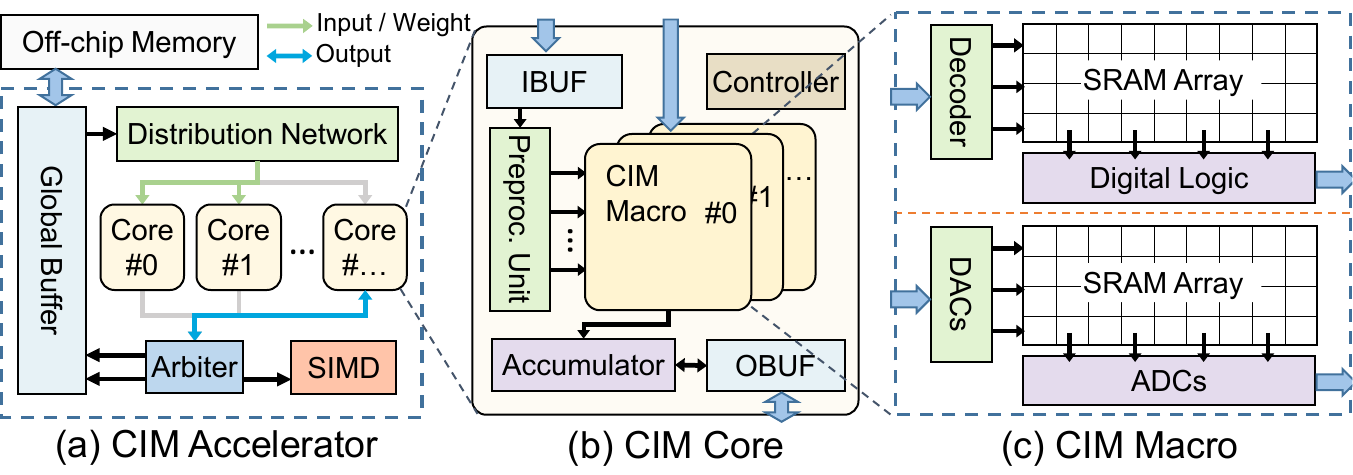}
  \caption{Hierarchical abstraction of the oriented CIM accelerator.}
  \label{fig:architecture}
\end{figure}

\subsection{Hardware Abstraction}
We illustrate the top-level accelerator architecture in Fig.~\ref{fig:architecture}(a), comprising a global buffer, a distribution network, multiple CIM cores, and a Single-Instruction Multiple-Data (SIMD) unit. 
The distribution network efficiently multicasts weights and features to the CIM cores for MVM operations. 
The SIMD unit is then responsible for post-processing, such as executing activation and pooling functions, and accumulating partial sums generated by individual cores.

The CIM core architecture shown in Fig.~\ref{fig:architecture}(b) features a configurable memory hierarchy of buffers and register files, which offers bypassable access to reduce latency and a double-buffered mode to overlap computation with communication.
However, this mode comes at the cost of halving the effective storage capacity compared to a single-buffered configuration.
In addition, the core datapath involves several non-bypassable processing stages, including a preprocessing unit for bit-serial input and an accumulator for partial sum aggregation.

\begin{figure}[t]
  \centering
  \includegraphics[width=0.85\columnwidth]{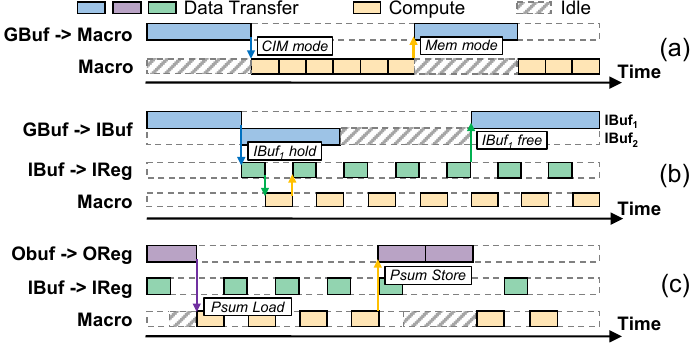}
  \caption{Representative data-transfer timelines illustrating (a) mode-switch stalls of the CIM macro, (b) pipeline stall due to throughput mismatch, and (c) operand-synchronization stalls.}
  \label{fig:pipeline}
\end{figure}

As depicted in Figure~\ref{fig:architecture}(c), the CIM macro stores weight data within its memory array. 
Input vectors are broadcast to the SRAM array, typically along the wordlines or bitlines, to perform MVM operations with weight data in either the analog or digital domain. 
CIM macros generally operate in two distinct modes: a standard \textit{Memory Mode} for read/write or weight update and a \textit{Compute Mode} for MVM execution. 
However, a critical limitation arises because both modes share peripheral circuits. 
This structural dependency prevents a conventional CIM array from performing computation and weight updates concurrently. 
Consequently, the resulting pipeline stalls create a severe performance bottleneck, an issue particularly acute for workloads that necessitate frequent weight reloading, as shown in Fig.~\ref{fig:pipeline}(a). 
While custom macro designs can circumvent this limitation, they often sacrifice versatility and introduce significant overhead in area, power, and design complexity.

\subsection{Data Transfer Analysis}
\label{sec3.2}

Previous analytical models for latency often oversimplify data transfer by assuming that latency equals the maximum access time across all memory hierarchy levels. 
However, this approach overlooks the complex interactions between operations executing within the accelerator. 
Even with double-buffering techniques, throughput mismatches between memory hierarchy levels can create pipeline stalls. 
As illustrated in Fig.~\ref{fig:pipeline}(b), the data transfer from the global buffer (GBuf) to the input buffer (IBuf) may be faster than the downstream processing at the input registers (IReg). 
Consequently, once $\text{IBuf}_2$ is also filled, the data transfer from the GBuf to the IBuf is forced into an idle state, awaiting $\text{IBuf}_1$ to become free.
This resulting pipeline stall significantly reduces the effective bandwidth of the datapath.

While the weight-stationary dataflow adopted by most CIM accelerators focuses on reducing weight data movement overhead, the transfer overhead of feature maps remains a significant factor in overall system performance.
As shown in Fig.~\ref{fig:pipeline}(c), partial sum (Psum) write-back operations cannot overlap with computation when the output register (OReg) employs single buffering, thereby lengthening the critical path. 
Furthermore, strict operand synchronization requirements create additional bottlenecks, where the delayed arrival of any operand stalls the entire pipeline, resulting in temporal underutilization of hardware resources.
\section{MIREDO Framework} 
\label{sec4} 

This section introduces MIREDO, an MIP-based framework designed for dataflow optimization in CIM accelerators, building on the architecture abstraction outlined in Section~\ref{sec3}.

\subsection{Framework Overview} 
\begin{figure}[t]
  \centering
  \includegraphics[width=0.77\columnwidth]{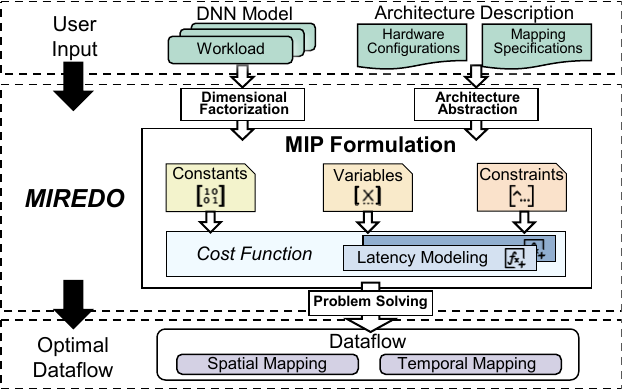}
  \caption{Overview of the proposed MIREDO framework.}
  \label{fig:overview}
\end{figure}

Dataflow optimization for DNN accelerators involves deriving optimal tiling factors and loop permutations for a given DNN workload, which can be represented via a loop nest format. 
To automate this complex process, we introduce MIREDO, a systematic framework that provides an adaptive approach to navigate the vast mapping space. 
The framework, shown in Fig.~\ref{fig:overview}, accepts a DNN model in ONNX format and a detailed architecture description as its primary inputs, producing a complete dataflow that explicitly defines the optimal temporal and spatial mapping.

The core of MIREDO is the mathematical formulation of the above optimization problem. 
The framework encodes the workload and hardware parameters as a set of constants $S_c$, while the dataflow mapping space is treated as a set of decision variables $S_v$.
A body of constraints links these constants and variables, restricting the search space by enforcing hardware limits and dataflow mapping specifications.
This structured representation constitutes an MIP problem in which the objective is to minimize a cost function $obj_{\text{cost}}$, encompassing not only overall latency but also memory access overhead.
Table~\ref{tab:mip_notations} summarizes the notations used in our model.
To enable an accurate cost evaluation, we further propose a novel analytical latency model that accurately accounts for pipeline stalls in Section~\ref{sec4.4}.
By solving the MIP problem formulated below:
\begin{equation}
    \min_{S_v} \,obj_{\text{cost}}(S_c, S_v),
\end{equation}
MIREDO produces a systematically optimized and highly efficient dataflow for the target CIM accelerator.

\begin{table}[t]
\small
\centering
\caption{MIREDO notation declarations.}
\renewcommand{\arraystretch}{0.95}
\resizebox{0.95\columnwidth}{!}{
\begin{tabular}{lll}
\toprule
\textbf{Index} & \textbf{Description} \\
\midrule
$d$ & Index for DNN tensor dimension.  \\
$f$ & Index for loop tiling factor. \\
$i$ & Index for temporal loop nest level. \\
$m$ & Index for memory hierarchy level.  \\
$u$ & Index for spatial unrolling axis. \\
$ \lambda $ & Index for operand type (input / weight / output) \\
\midrule

\textbf{Constant} & \multicolumn{2}{l}{\textbf{Description}} \\ 
\midrule
\( F \) & \multicolumn{2}{l}{Sets of tiling factors for each dimension.} \\
\( C \) & \multicolumn{2}{l}{Sets of constants defining the mapping specifications.} \\
\( H, Y \) & \multicolumn{2}{l}{Pre-enumerated sets for data sizes and loop bounds.} \\
\( \mu \) & \multicolumn{2}{l}{Weighting parameters for functions.} \\
\midrule

\textbf{Variable} & \multicolumn{2}{l}{\textbf{Description}} \\
\midrule
\( X \) & \multicolumn{2}{l}{Binary matrices representing spatial-temporal mapping.} \\
\( V \) & \multicolumn{2}{l}{One-hot encoded vectors for linear selection.} \\
\( \psi \) & \multicolumn{2}{l}{Binary indicators for operational states.} \\
\( T \) & \multicolumn{2}{l}{Transfer latency of the operands.} \\
\( P \) & \multicolumn{2}{l}{Processing latency of the operands.} \\
\( L \) & \multicolumn{2}{l}{Critical path latency at a loop level.} \\
\bottomrule
\end{tabular}
}
\label{tab:mip_notations}
\end{table}

\subsection{Constant Definitions} 
The constants in our model are derived from the hardware architecture and the workload.
A memory mapping matrix $C^M$ indicates whether each memory unit is dedicated to a single operand or shared among several operands. 
Furthermore, a key architectural characteristic for dataflow optimization is the available parallelism, which greatly influences computational efficiency and is implemented through spatial unrolling. 
This is captured by a binary matrix $C^X$ specifying which tensor dimensions can be unrolled onto these axes.
For instance, due to inherent hardware constraints, the CIM macro's wordline axis might only permit spatial unrolling at the output channel.
These formulations and other constants, such as memory bus width ($\mathcal{BW}$) and capacity ($\mathcal{CA}$), collectively define the rigid hardware resource constraints upon which dataflow optimization is performed.

DNN operators are typically modeled as loop nests, with bounds defined by dimensional parameters of the weights and feature maps. 
While prior work often decomposes these bounds into prime factors for allocation, this approach can lead to a combinatorial explosion in the search space due to the large number of factors $F$. 
To mitigate this complexity, we propose \textit{Flexible Factorization}: an algorithm that employs a greedy strategy to reduce the number of factors, as detailed in Alg.~\ref{alg:FlexibleFactorization}. 

The algorithm starts with a complete set of prime factors and iteratively merges pairs of factors. 
This process is controlled by two user-definable parameters: a minimum factor count ($k_{\min}$) and a relative loss threshold ($\alpha$). 
At each step, the algorithm chooses to merge the factor pair that incurs the minimum loss to the flexibility score, which is calculated by \textit{FlexScore} in Alg.~\ref{alg:FlexibleFactorization}.
The \textit{FlexScore} is a heuristic metric designed to quantify the mapping versatility of the factor set. 
Specifically, it enumerates all unique partitions of $F$ into $k$ disjoint subsets, where $k \in \{1, 2, 3\}$.
The score is computed as a weighted sum of the number of unique partitions, utilizing decreasing positive weights ($\mu^P$) to prioritize factor sets that offer greater partitioning flexibility. 
Merging factors diminishes this flexibility, consequently resulting in a lower score.
The merging process stops when the factor count reaches $k_{\min}$, or when the relative loss from the best available merge exceeds the threshold $\alpha$. 
This algorithm effectively reduces the search space complexity, providing a high-quality set of factors for the subsequent MIP optimization.

\SetKwProg{Fn}{Function}{:}{}

\begin{algorithm}[t]
\small
\footnotesize
\caption{Flexible Factorization}
\label{alg:FlexibleFactorization}
\KwIn{
    An integer $N > 1$, a minimum factor count $k_{\min}$, a relative loss threshold $\alpha \in [0, 1]$ \;
}
\KwOut{A factor list $F$ where $\prod F = N$.}
\BlankLine 

\Fn{FlexibleFactorization($N, \alpha, k_{\min}$)}{
    $F \leftarrow \text{PrimeFactors}(N)$\;
    \If{$\text{length}(F) \le k_{\min}$}{
        \KwRet $F$\;
    }
    $Score_{\text{full}} \leftarrow \text{FlexScore}(F)$\;
    \While{$\text{length}(F) > k_{\min}$}{
        $Score_{\text{base}} \leftarrow \text{FlexScore}(F)$\;
        $(\Delta_{\text{best}}, F_{\text{best}}) \leftarrow (\infty, \text{null})$\;
        \For{each pair $(f_i, f_j)$ in $F$}{
            $F_{\text{merged}} \leftarrow (F \setminus \{f_i, f_j\}) \cup \{f_i \cdot f_j\}$\;
            $Score_{\text{merged}} \leftarrow \text{FlexScore}(F_{\text{merged}})$\;
            $\Delta \leftarrow (Score_{\text{base}} - Score_{\text{merged}}) / Score_{\text{full}}$\;
            \If{$\Delta < \Delta_{\text{best}}$}{
                $\Delta_{\text{best}} \leftarrow  \Delta, F_{\text{best}} \leftarrow F_{\text{merged}}$\;
            }
        }
        \If{$\Delta_{\text{best}} > \alpha$}{
            \textbf{break} 
        }
        $F \leftarrow F_{\text{best}}$\;
    }
    \KwRet $F$\;
}
\Fn{FlexScore($F_{\text{input}}$)}{
    $P_1, P_2, P_3 \leftarrow \emptyset, \emptyset, \emptyset$\;
    $P_{\text{all}} \leftarrow \text{All partitions of } F_{\text{input}} \text{ into } k \in \{1, 2, 3\} \text{ subsets}$\;
    \For{each partition $\pi$ in $P_{\text{all}}$}{
        $k \leftarrow |\pi|$; 
        
        $prods \leftarrow \text{sorted tuple of products from subsets of } \pi$\;
        Add prods to $P_k$\;
    }
    \KwRet $\mu^P_1 \cdot |P_1| + \mu^P_2 \cdot |P_2| + \mu^P_3 \cdot |P_3|$\;
}
\end{algorithm}

\vspace{-5pt}

\subsection{Variables and Constraints} 
The formulation of the dataflow mapping space can be represented by a set of binary decision variables.
The fundamental mapping decision for each factor $F_{d,f}$ is its placement in either the temporal loop nest or a spatial unrolling dimension. 
This binary choice is captured by variables $X^L_{d,f,i}$ and $X^U_{d,f,u}$, respectively, governed by a uniqueness constraint:
\begin{equation}
    \sum_{i} X^L_{d,f,i} + \sum_{u} X^U_{d,f,u} = 1 \quad \forall d,f,
\end{equation} 
that ensures each factor is mapped exactly once.

To support flexible mapping strategies such as uneven mapping~\cite{mei2021zigzag}, we introduce a granular variable $X^M_{d,f,\lambda,m}$ that enables independent definition of loop blocks for each operand type ($\lambda$) at each memory level ($m$).
Loop blocks represent groups of loops assigned to the same memory level.
These variables are subject to the following constraints:
\begin{equation} \label{eq:Const-constraints}
\begin{aligned}
    X^M_{d,f,\lambda,m} &\leq C^M_{m,\lambda}, \; X^U_{d,f,u} \leq C^X_{u,d}, \\
    X^Z_{i,\lambda,m} &\geq X^M_{d,f,\lambda,m} \wedge X^L_{d,f,i}
\end{aligned}
\quad \forall d,f,\lambda,u,m,i.
\end{equation}
    
The variable $X^Z$ indicates whether memory level $m$ is mapped to loop $i$.
For the formulation of loop permutation, the number of available temporal loop slots is set equal to the total number of loop factors, with the indicator $\psi^L_i$ identifying which of these slots are active. Similarly, the indicator $\psi^U_{m,\lambda}$ specifies whether a memory level is utilized or bypassed for a particular operand.
\begin{gather}
    \psi^L_i = \sum_{d,f} X^L_{d,f,i},  \;
    \psi^U_{m,\lambda} = \bigvee_{d,f} X^M_{d,f,\lambda,m} \,
    \quad \forall i,\lambda,m.
\end{gather}
To formally define the data transfer path, we introduce $X^N$ as a binary variable indicating a direct transfer for operand $\lambda$ from $m$ to $m'$. 
The following constraints are introduced to link $X^N$ with the utilization status $\psi^U$, ensuring path uniqueness and preventing illegal bypasses:
\begin{equation}
    \sum_{m'\geq m+1}X^N_{m,m',\lambda} = \psi^U_{m,\lambda},\;
    1-\psi^U_{m_1,\lambda}+ X^N_{m,m_1,\lambda}\geq X^N_{m,m_2,\lambda}  .
\end{equation}
Here, $m$, $m_1$, and $m_2$ are memory level indices satisfying $m < m_1 < m_2$, where a larger index value $m$ denotes a memory level closer to the CIM macros.

To enforce hardware capacity constraints, the data size of each operand at every memory level must be precisely formulated.
The dimensional bound at a specific level is determined by multiplying all loop factors mapped to the current level and all levels below.
Since this multiplicative calculation is inherently non-linear, it is unsuitable for the MIP formulation.
The loop bound for each dimension ($B^S_{m,\lambda,d}$) is therefore calculated as a linear sum in the logarithmic domain:
\begin{equation} \label{eq:bound}
    B^S_{m,\lambda,d}=\sum_{f} \log(F_{d,f})\cdot(\sum_{m'=m}X^M_{d,f,\lambda,m'}+\sum_{u}X^U_{d,f,u} ), 
\end{equation}
where the summation over $u$ is performed for all indices satisfying the condition $C_u \geq m$.
Inspired by prior work~\cite{russo2023memory}, we circumvent the non-linear product by enumerating all valid data sizes and using a one-hot encoded vector ($V^S_{m,\lambda}$) to select a single candidate from the set $H_{m,\lambda}$.
Let $\mathcal{P}$ denote the operand precision, the data size is given by
\begin{equation} \label{eq:size}
    Size_{m,\lambda} = H_{m,\lambda}\cdot V^S_{m,\lambda} \cdot \psi^U_{m,\lambda} \cdot \mathcal{P}_{m,\lambda}  \quad \forall m,\lambda.
\end{equation}
A pre-calculated dimension vector $Y_{m,\lambda,d}$
holds the corresponding bound for each enumerated data size.
The constraint (\ref{eq:bound}) ensures that the loop bound derived from the mapping variables is consistent with the specific entry selected by the one-hot vector:
\begin{equation} 
    \log(Y_{m,\lambda,d}) \cdot V^S_{m,\lambda} = B^S_{m,\lambda,d} \quad \forall m,\lambda,d \; .
\end{equation}

The bit-serial and highly parallel nature of the CIM paradigm imposes unique demands on data transfer, necessitating flexible buffering strategies. 
Each memory level can operate in a standard single-buffered mode, which maximizes storage capacity but risks pipeline stalls due to mutually exclusive access. 
Alternatively, double-buffering can be employed to overlap data transfer with computation, at the expense of halving the effective storage capacity. 
To allow the dataflow optimizer to navigate this trade-off, the binary variable $\psi^{DM}$ is introduced to select the optimal mode, enabling the capacity constraint to be formally expressed as follows:
\begin{equation} \label{eq:capacity}
    \sum_{\lambda}  (1+\psi^{DM}_{m,\lambda}) \cdot Size_{m,\lambda}\leq \mathcal{CA}_m \quad \forall m.
\end{equation}

The size of the data tile transferred into a memory level is distinct from the data stored within it, due to data reuse and multicast opportunities.
We introduce a new dimensional bound ($B^T$) by modifying the formulation (\ref{eq:bound}) to exclude temporal loop factors mapped at the current memory level:
\begin{equation}
\hspace{-1mm}
    B^T_{m,\lambda,d}=\sum_f \log(F_{d,f})\cdot(\sum_{m'\geq m+1}X^M_{d,f,\lambda,m'}+\sum_{u}X^U_{d,f,u} ).
\end{equation}
A corresponding one-hot vector $V^T_{m,\lambda}$ links this new bound to the same constant $Y_{m,\lambda,d}$ via an analogous constraint.

\subsection{Performance Modeling} \label{sec4.4}
The total execution latency of a workload is determined by our loop-based analytical model, which recursively calculates the operand processing latency ($P_{i,\lambda}$) at each temporal loop level $i$.
Using the previously defined symbols, the cycle count for a transfer operation ($T_{i,\lambda}$) is defined by the following constraint:
\begin{equation} \label{eq:trans}
    X^Z_{i,\lambda,m}=1 \to T_{i,\lambda} \cdot \mathcal{BW}_m = H_{m,\lambda} \cdot V^T_{m,\lambda} \cdot \psi^U_{m,\lambda} \cdot \mathcal{P}_{m,\lambda}.
\end{equation}
The detailed formulations under all possible scenarios are listed in Table~\ref{tab:latency_expression}.
\begin{table}[t]
\small
\centering
    \caption{Classification of operand-processing latency for different buffering strategies and operand types (I: Input Feature, W: Weight, O: Output Feature).}
\renewcommand{\arraystretch}{1.5}

\resizebox{\columnwidth}{!}{
    \begin{tabular}{lll}
        \hline
        \textbf{Buf. Strategy} & \textbf{Operand} & \textbf{Latency expression ($L_{i,\lambda}$)} \\
        \hline
        Single&I / W&$L_i \cdot (N_i - 2) + 2*T_{i,\lambda} + P_{i+1,\lambda}$ \\
        Single&O&$L_i \cdot (N_i - 1) + 2*T_{i,\lambda} + P_{i+1,\lambda}$ \\
        Double&I / W&$\max\bigl\{L_i \cdot (N_i - 3) + 2*T_{i,\lambda} + \max\{T_{i,\lambda}, P_{i+1,\lambda}\}, T_{i,\lambda} \cdot N_i\bigr\}$ \\
        Double&O&$L_i \cdot (N_i - 2) + T_{i,\lambda} + \max\left(T_{i,\lambda}, L_i\right) + \max\left(T_{i,\lambda}, P_{i+1,\lambda}\right)$ \\
        (No Transfer)&(Any)&$L_i \cdot (N_i - 1) + P_{i+1,\lambda}$  \\
        \hline
        \end{tabular}
    }
    \label{tab:latency_expression}
\end{table}

The critical path latency $L_i$ is formulated as the maximum of two primary components: 
1) the cumulative latency of its nested inner loop, calculated as the product of its critical path $L_{i+1}$ and the loop count $N_{i+1}$; 
and 2) the combined latency of transfer and processing from operands at the current loop.
Since the product term is non-linear, MIREDO employs a one-hot vector $V^{F}$ to select a pre-enumerated bound, ensuring MIP compatibility.
This combination depends on the buffering strategy, captured by the binary indicator $\psi^{DL}_{i,\lambda}$:
\begin{equation} \label{eq:double_mem}
    \psi^{DL}_{i,\lambda} = \bigvee_{m,m'\geq m+1}(X^Z_{i,\lambda,m} \land X^N_{m,m',\lambda} \land \psi^{DM}_{m,\lambda}).
\end{equation}
This allows the model to select either a sequential summation ($T_{i,\lambda}+P_{i+1,\lambda}$) or an overlapped calculation ($\max\left(T_{i,\lambda}, P_{i+1,\lambda}\right)$) when double-buffering is enabled ($\psi^{DL}_{i,\lambda}$ = 1).

Operand-transfer latency is accounted for only in the innermost loop block of each memory level where the actual data transfer occurs. 
The model also incorporates the data-stationary scenario, which applies when an operand is reused at loop level $i$ and thus incurs no transfer latency.

For the innermost MVM operation within a nested loop, the latency is a constant ($L_\text{MVM}$) determined by the operand precision and the CIM macro design. 
The recursion boundary condition is therefore expressed as $L_{i_{max}+1} = P_{{i_{max}+1},\lambda} = L_\text{MVM}$.
To maintain model consistency, inactive temporal slots propagate latency from their inner loop, as expressed by the relation:
\begin{equation} \label{eq:inactiveSlot}
    \psi_{i}^{L}=0\rightarrow(L_{i},P_{i,\lambda})=(L_{i+1},P_{i+1,\lambda})  \quad \forall i,\lambda.
\end{equation}

Based on the recursive components, the final optimization objective is formulated as:
\begin{equation} \label{eq:objective}
    obj_{\text{cost}} = \mu^C_1 \cdot \max_\lambda(P_{0,\lambda}) - \mu^C_2 \cdot \sum_{m,\lambda} m * Size_{m,\lambda}.
\end{equation}
This cost function balances two competing goals via weighting parameters $\mu^C$: minimizing total latency and maximizing data locality.
The latter goal incentivizes storing data at lower memory hierarchy levels closer to the CIM macros to reduce data movement overhead.
Minimizing this objective function yields a dataflow optimized for both execution performance and resource efficiency.

\section{Evaluation Results}

\subsection{Experiment Setup}

\begin{table}[t]
\small
\centering
    \caption{Hardware component configurations.}
\begin{tabular}{|c|c|c|}
    \hline
    \textbf{Component} &  \multicolumn{2}{c|}{\textbf{Configuration}}   \\ 
    \hline
    CIM Macro & Array & $128 \times 32$ \\ \hline
    \multirow{2}{*}{Local Buffer} & Capacity & 256 KB \\
    & Bus Width & 128 bit \\ \hline
    Core & Number & 8\\ \hline
    \multirow{2}{*}{Global Buffer} & Capacity & 8 KB \\
    & Bus Width & 256 bit \\ \hline
    
    Off-chip Memory & Bus Width & 64 bit \\ \hline
\end{tabular}
    \label{tab:Components}
\end{table}

To ensure a fair and accurate evaluation of MIREDO, we developed a custom simulator based on the architecture from Section~\ref{sec3}. 
Latency and power of memory are modeled using PCACTI~\cite{shafaei2014fincacti}, and the statistics of other logic components are derived from prior works~\cite{sun2023analog}.
The detailed configurations of these hardware components are summarized in Table~\ref{tab:Components}. 
We employ Gurobi~\cite{gurobi}, a general-purpose optimizer for MIP and other constrained programming, as the solver. 
By automatically identifying convolutional layers from the DNN model, MIREDO defines the constants, variables, constraints, and objective functions before invoking the solver.
We equip a 2GHz Intel Xeon Gold 6330 CPU, with the time for solving a single layer capped at 5 minutes.

To evaluate MIREDO, we benchmark its performance against two representative dataflow schemes:
1) A heuristic search inspired by the ZigZag framework~\cite{sun2023analog} for its support of uneven mappings;
2) A conventional Weight-Stationary (WS) dataflow derived by imposing additional constraints within our own MIP formulation.
This setup provides a comparison against both an empirical strategy and a heuristic optimization approach.
Our baseline workload utilizes ResNet-18 inference on ImageNet~\cite{deng2009imagenet}, with weight and activation quantized to INT8 format.
Subsequently, we extended our analysis to encompass various networks and architectures.
\begin{figure}[t]
  \centering
  \includegraphics[width=0.95\columnwidth]{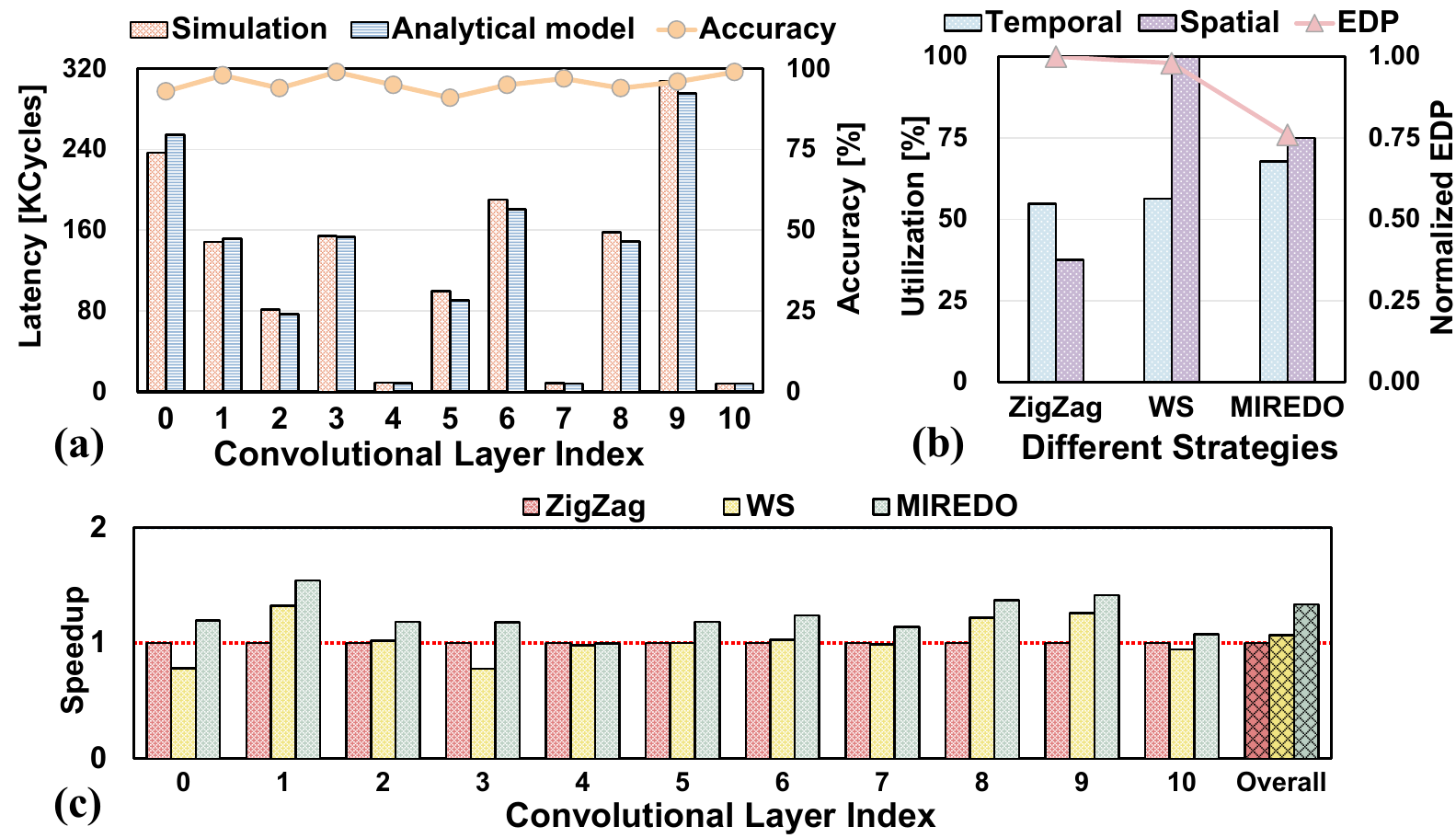}
  \caption{MIREDO performance evaluation. (a) Analytical model accuracy validation. (b) Utilization and EDP comparison. (c) Per-layer and overall speedup comparison. }
  \label{fig:effective}
\end{figure}

\subsection{Performance Evaluation}
\textbf{Accuracy.} The accuracy of our analytical model was validated by comparing its latency predictions against results from a detailed hardware simulation. 
As illustrated in Fig.~\ref{fig:effective}(a), the model achieves an average accuracy of $95.5\%$ across all evaluated model layers.

\textbf{Latency.} 
We further compare the latency of different dataflow strategies per layer, as depicted in Fig.~\ref{fig:effective}(c). 
The heuristic scheme from ZigZag and the constrained weight-stationary dataflow exhibit comparable overall latency, yet their effectiveness varies across different layers. 
In contrast, MIREDO achieves a consistent speedup across most layers.

\textbf{Efficiency.} 
Fig.~\ref{fig:effective}(b) illustrates the trade-off between macro utilization and overall system efficiency in dataflow optimization for a representative layer. Compared to a baseline with limited optimization, the weight-stationary method maximizes spatial utilization to improve weight reuse, but restricts the search space and leads to a suboptimal solution. By jointly optimizing latency and resource allocation, MIREDO explores a considerably larger mapping space and consequently improves system-level performance.

\begin{figure}[t]
  \centering
  \includegraphics[width=0.95\columnwidth]{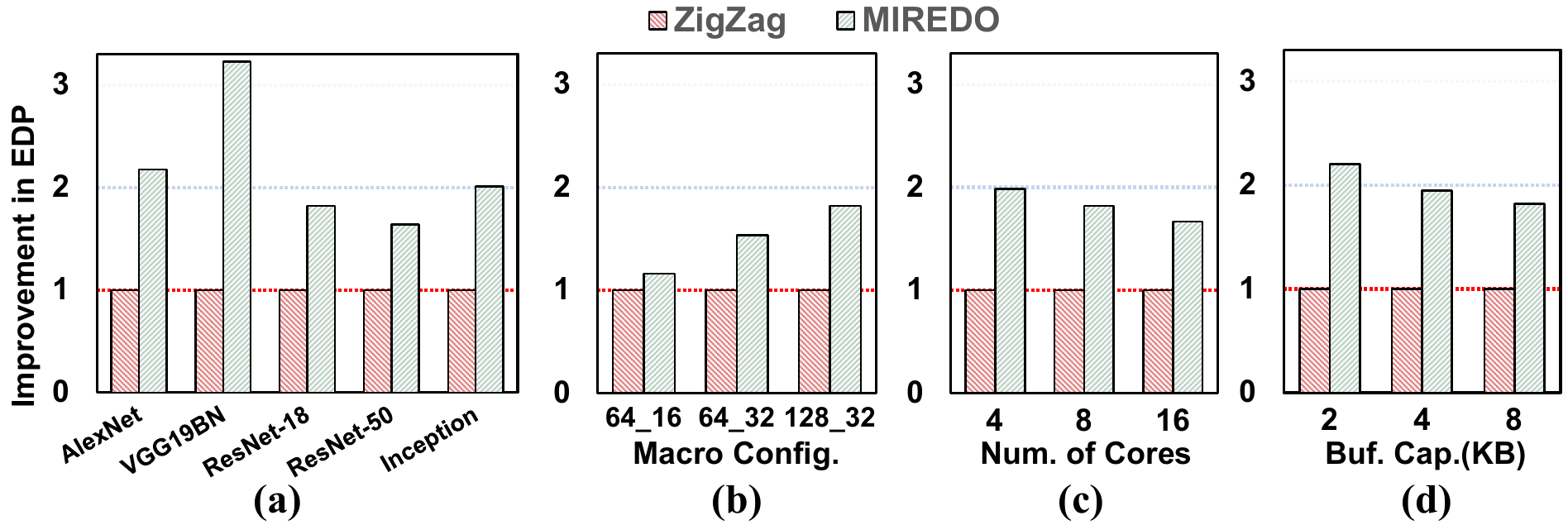}
  \caption{Performance comparison across various DNN models and hardware configurations.}
  \label{fig:Adaptability}
\end{figure}
\subsection{Adaptability and Robustness}

We conduct extensive evaluations to validate the adaptability and robustness of MIREDO across diverse workloads and hardware configurations.
Our evaluation first assesses MIREDO's adaptability against the ZigZag-based heuristic across representative DNN models. 
As shown in Fig.~\ref{fig:Adaptability}(a), our approach consistently achieves significant EDP reductions, ranging from $1.6\times$ to $3.2\times$ over the baseline.  
These results demonstrate MIREDO's ability to generate optimized dataflows tailored to different workloads.

To assess robustness, we analyze performance across various hardware configurations, including different macro configurations, core counts, and buffer capacities.
As illustrated in figures~\ref{fig:Adaptability}(b)-(d), MIREDO achieves significant improvements in EDP reduction compared to the baseline across various hardware configurations.
This highlights the value of systematic optimization in memory-constrained environments, where conventional approaches struggle to balance efficiency with severe memory limits.

\section{Conclusions}

In this work, we introduce MIREDO, a novel framework utilizing MIP to optimize dataflows for executing DNN workloads on CIM accelerators. 
Through a hierarchical hardware abstraction, MIREDO systematically models the complex mapping processes and hardware constraints, allowing accurate estimation of execution latency. 
This structured approach enables MIREDO to efficiently determine the optimal dataflow configuration in a single iteration, significantly improving CIM accelerator efficiency.
Experimental results show that MIREDO effectively reduces the EDP by up to $3.2\times$, highlighting its adaptability to diverse workloads and hardware constraints. 
Moreover, the flexible MIP formulation ensures that MIREDO is highly scalable and easily extendable to various DNN workloads and CIM architectures.

\clearpage

{
\normalsize
\bibliographystyle{IEEEtran}
\bibliography{ref}
}

\end{document}